\documentclass[%
 amsmath,amssymb,
 article,
]{revtex4-2}

\usepackage{graphicx}
\usepackage{dcolumn}
\usepackage{bm}
\usepackage{color}
\usepackage{soul}

\begin{document}


\title{Selectable diffusion direction with topologically protected edge modes}

\author{Keita Funayama}
\email{funayama@mosk.tytlabs.co.jp}

\author{Atsushi Miura}
\author{Hiroya Tanaka}
\affiliation{Toyota Central R\&D Labs., Inc., Nagakute 480-1192, Japan}

\author{Jun Hirotani}%
\affiliation{Department of Micro Engineering, Kyoto University, Kyoto 615-8540, Japan}%


\date{\today}

\begin{abstract}
Topological insulators provide great potentials to control diffusion phenomena as well as waves.
Here, we show that the direction of thermal diffusion can be selected by the contributions of the topologically protected edge modes via the quantum spin Hall effect in a honeycomb-shaped structure. 
We demonstrate that when we set our structure to the temperature corresponding to the type of edge mode, the direction of thermal diffusion can be tuned.
Moreover, this diffusion system is found to be immune to defects owing to the robustness of topological states.
Our work points to exciting new avenues for controlling diffusion phenomena.

\end{abstract}

\maketitle



\section{Introduction}

There is now significant interest in exploiting topological properties in a wide variety of physical wave systems as electromagnetic~\cite{Zhao2018,Li2018,Wang2009,Hasan2010,Qi2011,Lee2018}, acoustic~\cite{Darabi2020,Lee2019,Zhang2018}, and mechanical systems~\cite{Liu2019,Yu2018,Cha2018,Vila2017,Wang2020,Matlack2018,Funayama2022}.
One of the characteristic topological phenomena is the emergence of robust edge modes.
The methodology based on the Su-Schrieffer-Heeger (SSH) model is the most common method for the emergence of edge modes~\cite{Peng2017,Su1979}.
This has allowed the wave control with topological insulators in one-dimensional classical systems~\cite{Lan2021,Lin2021,Engelhardt2017,Tian2022}.
In addition, the quantum spin Hall effect (QSHE) has attracted significant attention owing to its higher degree of freedom (DOF)
than that of the SSH model.
The QSHE-based topological states can control wave systems in higher dimensions, thereby enabling localization~\cite{Zangeneh2019,Fan2019,Schindler2018} and one-way wave propagation~\cite{Wu2015,Zhang2017,Davis2021,Yu2019}.

Recent theoretical and experimental studies have demonstrated that topological edge modes can be applied to diffusion systems using the SSH model~\cite{Hu2022,Qi2022,Yoshida2021,Xu2022}.
Such edge modes enable temperature localization and robust thermal decay.
The pioneering studies implicate the potential for the diffusion control at higher dimensions using alternative topological states.
However, topological states with higher DOF have not been investigated sufficiently yet in diffusion systems.

In this paper, we demonstrate that QSHE-based topologically protected edge modes appear in a thermal diffusion system.
Our structure consists of honeycomb-shaped unit cells with topological and ordinary states.
We consider that the heat transfer in our structure at the topological edge modes appears around the boundary between the topological and ordinary states. 
From numerical and analytical studies, we show that the temperature corresponding to the edge modes provides the directional heat transfer.
The honeycomb-shaped unit cells function as a heat-transfer path while maintaining the edge modes in our structure.
Consequently, the diffusion direction can be selected based on the type of excited edge mode.
We also verified a well-known unique characteristic of topological edge modes, which is that they are immune to defects.
Our results indicate that the use of QSHE-based topological edge modes has the potential to control heat transfer in any direction. 
Generally, our work should motivate systematic studies to apply topological properties to all diffusion phenomena.

\section{Results}
\subsection{Design of a topological diffusion system}
\begin{figure}[h]
	\centering
	\includegraphics[width=0.48\textwidth]{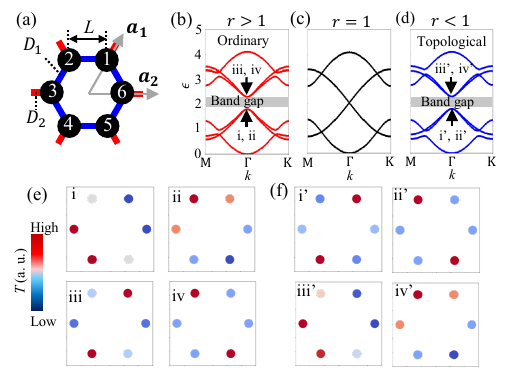}
	\caption{
        (a) Schematic of the honeycomb-shaped unit cell.
        Each of the nearest neighboring sites (black) is connected by the beams with the effective diffusivity $D_1$ (blue).
        The neighboring unit cells are connected by the beams with the effective diffusivity $D_2$ (red). 
        The effective diffusivities $D_1$ and $D_2$ are normalized to $L^{-2}$, where $L=35$~mm is the length of the beams.
        (b), (c), (d) Spectra of the eigenvalues for the infinite periodic honeycomb lattice with (b) $r>1$ ($D_1=0.765$ s$^{-1}$,  $D_2=0.518$ s$^{-1})$, (c) $r=1$ ($D_1=D_2=0.68$ s$^{-1}$), and (d) $r<1$ ($D_1=0.6$ s$^{-1}$, $D_2=0.85$ s$^{-1}$).
        Each of the doubly degenerated modes in (b) and (d) are labeled as i-iv and i'-iv'. 
        (e), (f) Site temperature corresponding to the eigenfunctions in the unit cell at modes (e) i-iv and (f) i'-iv'.
        }
	\label{fig:figure1}
\end{figure}

Our structure consists of periodically aligned honeycomb-shaped unit cells, as illustrated in Fig.~\ref{fig:figure1}(a). 
The unit cell has six circle sites. 
The nearest neighboring sites and unit cells are connected by fine beams with the effective diffusivities $D_1$ and $D_2$. 
As a result, the equation of thermal diffusion in a unit cell is expressed by the $6\times6$ effective diffusivity matrix,
\begin{align}
\label{eq:1a}
\frac{\partial}{\partial t}
\begin{bmatrix}
T_{1}  \\
\vdots \\
T_{6}
\end{bmatrix}
&=
\begin{bmatrix}
Q & P_{1} \\
P_{2} & Q
\end{bmatrix}
\begin{bmatrix}
T_{1}  \\
\vdots \\
T_{6}
\end{bmatrix},
\end{align}
where
\begin{align}
\label{eq:1b}
Q
&=
\begin{bmatrix}
-\left(2D_{1}+D_{2}\right) & 0 & 0\\
0 & \ddots & \vdots  \\
0 & \cdots & -\left(2D_{1}+D_{2}\right)
\end{bmatrix},
\end{align}
\begin{align}
\label{eq:1c}
P_{1\left(2\right)}
&=
\begin{bmatrix}
D_{2}e^{ik\left(\bm{a}_{1\left(2\right)}-\bm{a}_{2\left(1\right)}\right)} & D_{1} & D_{1}  \\
D_{1} & D_{1} & D_{2}e^{ik\bm{a}_{1\left(2\right)}} \\
D_{1} & D_{2}e^{-ik\bm{a}_{2\left(1\right)}} & D_{1} \\
\end{bmatrix}.
\end{align}
Here, eigenfunctions, $T_1, \ldots,$ and $ T_6$, denote the temperature at each site as numbered in Fig.~\ref{fig:figure1}(a), $k$ is the wavenumber, and $\bm{a}_{1(2)}$ is the unit vector as shown in Fig.~\ref{fig:figure1}(a).

We design the topological and ordinary unit cells by adjusting the ratio $r=D_1/D_2$ of the two effective diffusivities.
The QSHE-based topological phase transition can be controlled via $r$~\cite{Li2018,Funayama2022}.
Diagonalizing the effective diffusivity matrix in Eq.~\eqref{eq:1a}, we obtain the spectrum of the eigenvalues $\epsilon$.
Figures~\ref{fig:figure1}(b)-(d) show the spectra for an infinite periodic honeycomb lattice with $r>1$, $r=1$, and $r<1$.
When all diffusivities in the unit cell have the same value (i.e., $r=1$), we observe the Dirac cone at $\Gamma$ point [Fig.~\ref{fig:figure1}(c)].
For $r>1$ and $r<1$, the spectra have a bandgap and two doubly degenerated modes [Figs.~\ref{fig:figure1}(b) and \ref{fig:figure1}(d)].

To identify the state of the structure with $r>1$ and $r<1$, we confirm the eigenfunctions of each doubly degenerated mode specified as i-iv in Fig.~\ref{fig:figure1}(b) and i'-iv' in Fig.~\ref{fig:figure1}(d).
The eigenfunctions in these modes correspond to the site temperature in the unit cell.
Figure~\ref{fig:figure1}(e) shows the temperature distributions of modes i-iv for $r>1$.
We observe the dipole $p_{x}$ and $p_{y}$ modes at modes i and ii, and quadrupole $d_{xy}$ and $d_{x^2-y^2}$ modes at modes iii and iv, respectively.
Thus, the lowly and highly polarized modes appear bellow and above the band gap, respectively.
This result indicates that the structure with $r>1$ is the ordinary state.
In contrast, in Fig.~\ref{fig:figure1}(f), for $r<1$, modes i' and ii' (iii' and iv') show the quadrupole (dipole) modes. 
Such an inversion of the order between the dipole and quadrupole modes signifies that the structure with $r<1$ is the nontrivial topological state.

\begin{figure}[h!]
	\centering
	\includegraphics[width=0.7\textwidth]{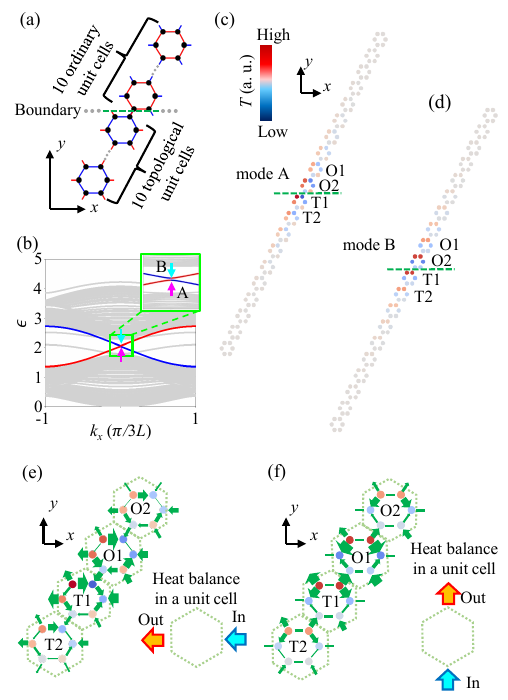}
        \caption{(a) Schematic of the supercell with a boundary between the topological and ordinary unit cells. 
        The supercell consists of 10 units of topological and ordinary states.
        (b) Band diagram of the supercell. 
        The blue and red lines denote the topological edge modes.
        Modes A and B are indicated by magenta and cyan arrows, respectively.
        The inset shows the enlarged view of the green square.
        (c), (d) Temperature distributions in the supercell at modes (c) A and (d) B. 
        (e), (f) Heat transfer between the sites for the unit cells O2, O1, T1, and T2 at modes (e) A and (f) B.  
        Green arrows denote the flow direction, and their widths imply the transferred heat quantity between the nearest neighboring sites.
        Right bottom panel in (e) and (f) shows the heat balance in each unit cell.
        }
	\label{fig:figure2}
\end{figure}

To demonstrate the QSHE-based topologically protected edge modes in the thermal diffusion system, we consider a supercell with a boundary between the topological and ordinary states, as shown in Fig.~\ref{fig:figure2}(a).
Figure~\ref{fig:figure2}(b) shows the spectra of the supercell.
The red and blue lines show the topological edge modes with different polarizations in the unit cells.
The gray lines indicate the bulk modes.
We focus on the two band-gap-crossing edge modes at $k_x=0$, indicated by the magenta and cyan arrows in Fig.~\ref{fig:figure2}(b) and are specified as modes A and B, respectively.

Figures~\ref{fig:figure2}(c) and \ref{fig:figure2}(d) depict the temperature distributions in modes A and B in the supercell.
Both temperature distributions are localized around the boundary.
The temperatures in modes A and B correspond to the eigenfunction amplitudes obtained by solving the eigenvalue equation, which consists of thermal diffusion equations for all 120 sites in the supercell.
Such edge modes have great potential for controlling diffusion phenomena.
Indeed, in wave systems, topologically protected edge modes have intriguing characteristics, such as field localization and unidirectional wave propagation~\cite{Yu2018,Cha2018,Li2018,Fan2019,Funayama2022,Lee2019}.
 
Based on the eigenfunction amplitudes on each site, we analyze heat transfer in modes A and B.
Figures~\ref{fig:figure2}(e) and \ref{fig:figure2}(f) visualize the diffusion direction and quantity of heat transferred between the sites in the two ordinary and topological unit cells (specified as O2, O1, T1, and T2).
The green arrows denote the flow direction, and their widths are the transferred heat quantity between the nearest neighboring sites.

Focusing on the heat balance in the individual unit cells, we find that modes A and B have a unique diffusion direction. 
We calculate the heat balances by summarizing the heat quantities along the $x$- and $y$-axes [right panels in Figs.~\ref{fig:figure2}(e) and \ref{fig:figure2}(f)].
In Fig.~\ref{fig:figure2}(e), the heat balance indicates the thermal flow along the $x$-axis through the unit cell; there is no heat balance along the $y$-axis.
Thus, mode A transfers heat only along the direction parallel to the boundary.
On the other hand, as illustrated in Fig.~\ref{fig:figure2}(f), mode B provides thermal flow only along the $y$-axis.
Therefore, our structure has the potential to select the diffusion direction by exciting the different edge modes.



\subsection{Edge states and heat transport}
\begin{figure}[b!]
	\centering
	\includegraphics[width=0.7\textwidth]{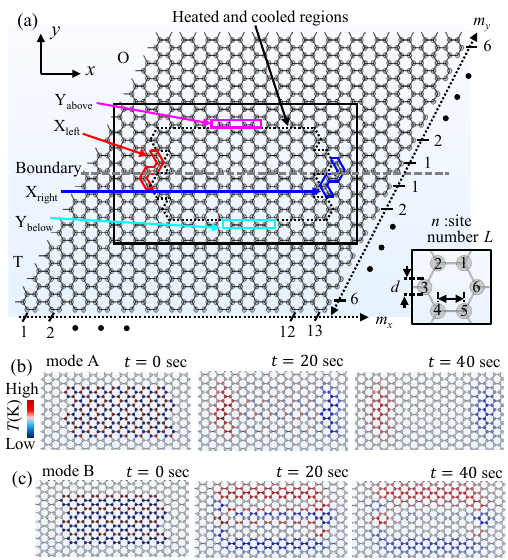}
        \caption{(a) Structure with the boundary between the topological and ordinary unit cells. 
        The edge modes are excited in the region enclosed by the broken black line.  
        Initial temperature is $T_{\mathrm{0}}=293.15$ K in the area except the enclosed region. 
        Regions X$_{\mathrm{left}}$ (red line), X$_{\mathrm{right}}$ (blue line), Y$_{\mathrm{above}}$ (magenta line), and Y$_{\mathrm{below}}$ (cyan line) are the nearest neighboring sites from the heated and cooled regions.
        The thickness of the simulated model is 10 mm, the diameter of the site is $d=20$ mm, and the distance between the nearest neighboring sites is the same as the value for the model in Fig.~\ref{fig:figure1} ($L=35$ mm). 
        (b), (c) Snapshots of thermal diffusion at $t=0$, $20$ s, and $40$ s for modes (b) A and (c) B in the area enclosed by the solid black square in (a).
        }
	\label{fig:figure3}
\end{figure}

To verify our theoretical prediction for selecting the diffusion direction, we compute the time evolution of the temperature in modes A and B using COMSOL Multiphysics. 
Figure~\ref{fig:figure3}(a) shows the model used in the calculation.
Each of the half structures in Fig.~\ref{fig:figure3}(a) consists of topological (T) or ordinary (O) states.
For the numerical investigation, we convert the eigenfunction to temperature $T_{s, \bm{v}}=T_{0}+\alpha T_{s, \bm{v}}^{\mathrm{A}\left(\mathrm{B}\right)}$ in Kelvin.
Here, $T_{\mathrm{0}}=293.15$ K is the reference temperature, $\alpha=100$ is the amplification coefficient, $T_{s, \bm{v}}^{\mathrm{A}\left(\mathrm{B}\right)}$ is the eigenfunction of mode A (B) at location $\bm{v}=[m_x, m_y ,n ]$ of the $n$-th site $(n\in\{1,\cdots,6\})$ in the $m_x$-th and $m_y$-th cells ($m_x \in \{1,\cdots,13\}$ and $m_y \in \{1,\cdots,6\}$), and $s\in\{\mathrm{O},\mathrm{T}\}$ is the ordinary
and topological sites.
To excite the edge modes to the system, we apply modes A or B to the unit cells of $m_y=1$ and $m_y=2$, which are enclosed by the broken black line in Fig.~\ref{fig:figure3}(a).
The other sites are set as $T_{\mathrm{0}}$ because the eigenfunction amplitudes are negligibly small in the unit cells of $m_y\geq3$,
due to the strong localization of the field around the boundary at the edge modes.
In fact, we observe that the temperature at the site of $m_y=3$ ($T_{s,m_x,3,n}$) is 30\% or less of $m_y=$ 1 ($T_{s,m_x,1,n}$) for all $m_x$ and $n$ values.

Figures~\ref{fig:figure3}(b) and \ref{fig:figure3}(c) show the numerical time evolution of the temperature distributions for modes A and B in the region enclosed by the solid black square in Fig.~\ref{fig:figure3}(a).
When exciting mode A, we observe high and low temperatures at $t = 40$ s at left and right edges  [regions X$_{\mathrm{left}}$ and X$_{\mathrm{right}}$ in Fig.~\ref{fig:figure3}(a)], see Fig.~\ref{fig:figure3}(b).
Interestingly, in the topological diffusion system, another edge mode enables heat transfer along the direction perpendicular to the boundary, unlike the topological wave systems.
When exciting mode B, we indeed observe high and low temperatures at above and below edges [regions Y$_{\mathrm{above}}$ and Y$_{\mathrm{below}}$ in Fig.~\ref{fig:figure3}(a)], see Fig.~\ref{fig:figure3}(c).
Thus, modes A and B realizes thermal diffusion only along the $x$- and $y$-axis, respectively, as predicted in Fig.~\ref{fig:figure2}(e) and \ref{fig:figure2}(f).

\begin{figure}[t]
	\centering
	\includegraphics[width=0.6\textwidth]{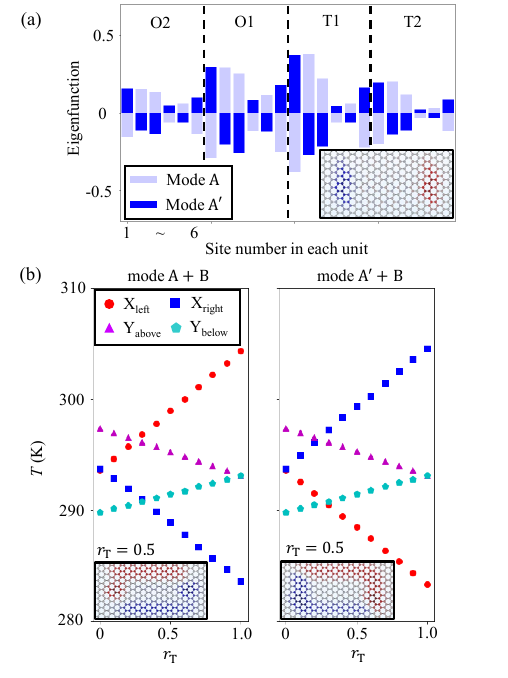}
	\caption{(a) Eigenfunctions of modes A (light blue) and A$^{\prime}$ (dark blue) on the sites in the unit cells specified as O2, O1, T1, and T2 in Fig.~\ref{fig:figure2}.
    The inset shows the temperature distribution at $t=40$ s, when mode A$^{\prime}$ is excited  to the structure in Fig.~\ref{fig:figure3}(a). 
    (b) Temperature in regions X$_{\mathrm{left}}$(red), X$_{\mathrm{right}}$(blue), Y$_{\mathrm{above}}$(magenta), and Y$_{\mathrm{below}}$(cyan) at $t=40$ s as the function of $r_{\mathrm{T}}$ when modes A+B (left) and A$^{\prime}$+B (right) are applied to the structure in Fig.~\ref{fig:figure3}(a).
    In the calculations for the inset of (a) and (b), we use the same simulation settings as that in Fig.~\ref{fig:figure3} except for the initial temperature determined by the applied edge modes.
    }
	\label{fig:figure4}
\end{figure}

Considering the antiphase of the edge modes, we can further select the diffusion direction.
As the phase of the fields is time-invariant in diffusion systems, inphase mode A (B) and antiphase mode A$^{\prime}$ (B$^{\prime}$) provides symmetric temperature distributions about the $y$- ($x$-)axis. 
When the temperature distribution of the inphase and antiphase modes coincides with the symmetry of the structure, the eigenequation contains antiphase modes as the eigenvalue solution.
Because our structure is axisymmetric about the $y$-axis, mode A$^{\prime}$ can appear as an edge mode.
Figure~\ref{fig:figure4}(a) shows the eigenfunctions in cells O1, O2, T1, and T2 of the supercell in mode A (light blue) and A$^{\prime}$ (dark blue).
Mode A$^{\prime}$ has an eigenvalue at $k_{x}=0.02$, as shown by the blue line in Fig.~\ref{fig:figure2}(b).
The eigenfunction of mode A$^{\prime}$ exhibits an inverted distribution of mode A at site O2 (also at T1, O2, and T2).
Thus, mode A and A$^{\prime}$ are mutually antiphase 
and this mode property is matched to the symmetry of the structure. 

Importantly, the inverted initial field about the $y$-axis affects the diffusion direction.
The inset in Fig.~\ref{fig:figure4}(a) shows the snapshot of the temperature distribution at $t = 40$ s when mode A$^{\prime}$ is excited to the structure in Fig.~\ref{fig:figure3}(a). 
We clearly see high and low temperatures in regions X$_{\mathrm{right}}$ and X$_{\mathrm{left}}$, respectively. 
This proves that mode A$^{\prime}$ inverts the diffusion direction of mode A.
However, mode B$^{\prime}$ does not appear in the solutions of the eigenequation based on our supercell because our structure is asymmetry about the $x$-axis.
Thus, we can change the diffusion direction by selecting the initial temperature of the sites based on the three different modes (A, A$^{\prime}$, and B) in our structure.

Furthermore, our structure enables to control the diffusion direction owing to the incoherence of the edge modes.
Specifically, the mutually incoherent modes A (A$^{\prime}$) and B can be linearly combined.
Using the ratio $r_{\mathrm{T}}$ of modes A (A$^{\prime}$) and B, we can excite our structure in the combined mode as 
\begin{align}
\label{linear comb}
T_{s,\bm{v}}^{\mathrm{AB}}= r_{\mathrm{T}}T_{s,\bm{v}}^{\mathrm{A}}+(1- r_{\mathrm{T}})T_{s,\bm{v}}^{\mathrm{B}}, \\
\label{linear comb AdB}
T_{s,\bm{v}}^{\mathrm{A'B}}= r_{\mathrm{T}}T_{s,\bm{v}}^{\mathrm{A'}}+(1- r_{\mathrm{T}})T_{s,\bm{v}}^{\mathrm{B}},
\end{align}
where $T_{s,\bm{v}}^{\mathrm{A'}}$ is the eigenfunction of mode A$^{\prime}$ and $0 \leq r_{\mathrm{T}} \leq 1$.
Figure~\ref{fig:figure4}(b) shows the temperature dependence of $ r_{\mathrm{T}}$ in regions X$_{\mathrm{left}}$, X$_{\mathrm{right}}$, Y$_{\mathrm{above}}$, and Y$_{\mathrm{below}}$ in Fig.~\ref{fig:figure3}(a) at $t=40$ s.
Depending on the type of the combined mode, the temperature in the four regions linearly increases and decreases with an increase in $r_{\mathrm{T}}$.
In particular, when $r_{\mathrm{T}}=0.5$, we obtain a symmetric temperature distribution about the $y$-axis, see the inset of Fig.~\ref{fig:figure4}(b).
The results here indicate the potential for designing arbitral temperature distribution and multidirectional heat transfer.

\subsection{Robustness of the edge modes}
\begin{figure*}[t]
	\centering
	\includegraphics[width=1\textwidth]{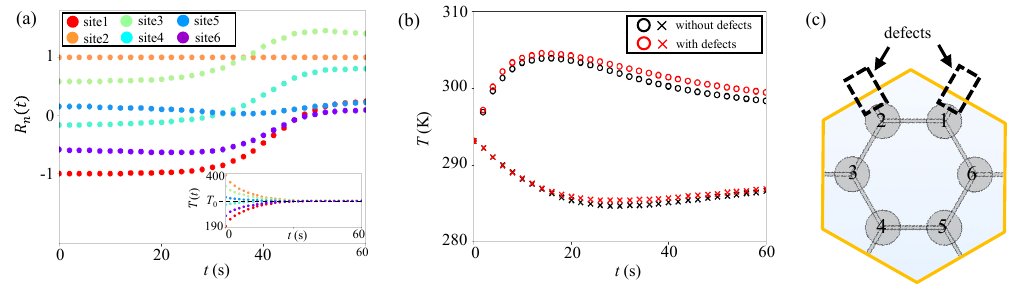}
	\caption{(a) Time dependency of the relative temperature $R_{n}\left(t\right)$ on the six sites in the topological unit cell of $(m_x,m_y)=(7,1)$ at mode A.
    The inset shows the temperatures in time on each site.
    (b) Time dependency of the temperatures in regions X$_{\mathrm{left}}$ (circles) and X$_{\mathrm{right}}$ (crosses). Red and black symbols show the results with and without the defect, respectively. 
    (c) Schematic of the unit cell including the defects. Sites 1 and 2 lack beams.
    We use the same simulation settings as that in Fig. 3 except for the defects in (c).
    }
	\label{fig:figure5}
\end{figure*}

The edge modes gradually collapse with time after those are applied to the structure.
As directional heat transfer of the unit cells continues as long as the edge modes are  maintained, the temporal robustness of the edge modes is crucial for designing topological diffusion systems.
Focusing mode A, we evaluate the temperature variation with time for the topological unit cell.
Figure~\ref{fig:figure5}(a) shows the relative temperature $R_{n}(t)= \tilde{T}_{\mathrm{T},n}(t)/\tilde{T}_{\mathrm{T},2}(t)$, considering the unbiased temperature $\tilde{T}_{\mathrm{T},n}(t) = (T_{\mathrm{T},7,1,n}(t)-T_0)/\alpha$ of the unit cell near the boundary, $(m_x,m_y)=(7,1)$ in Fig.~\ref{fig:figure3}(a).
For $t < 20$ s, $R_{n}(t)$ on each site has a constant value, that is, each temperature uniformly decays at the topologically protected decay rate.
For $t \geq 20$ s, mode A collapses because the temperatures of the unit cell almost converge to $T_{0}$, see inset of Fig.~\ref{fig:figure5}(a); thus mode A cannot be maintained.
The temporal mode robustness contributes to the duration of the directional heat transfer.
Figure~\ref{fig:figure5}(b) shows the temperature in regions X$_{\mathrm{left}}$ (circles) and X$_{\mathrm{right}}$ (crosses) as a function of time. 
For $t < 20$ s, in region X$_{\mathrm{left}}$ (X$_{\mathrm{right}}$), the temperature increases (decreases) with time due to the directional heat transfer provided by mode A.
For $t \geq 20$ s, the temperature in X$_{\mathrm{left}}$ (X$_{\mathrm{right}}$) decays (saturates) because the edge mode plays a minor role in the structure.

Finally, we evaluate a unique characteristic of the topologically protected edge modes, which is their immunity to defects.
To introduce the defects into our structure, we remove the beams from sites 1 and 2, see Fig.~\ref{fig:figure5}(c), for two unit cells $(s,m_x,m_y)=(\mathrm{T},6,1)$ and $(\mathrm{T},8,1)$.
The time dependence of the temperature in regions X$_{\mathrm{left}}$ and X$_{\mathrm{right}}$ is plotted with red symbols in Fig.~\ref{fig:figure5}(b). 
We observe small variations in the temperatures with and without defects.
Hence, the edge states are robust against disorders due to topological protection as well as other topological systems.

\section{Discussion}
We have shown that QSHE-based topologically protected edge modes appear in the diffusion system consisting of the honeycomb-shaped structure.
While previous topological edge modes in diffusion systems could only design the decay rate~\cite{Hu2022,Qi2022,Yoshida2021}, the QSHE based edge states in our structure realize that the diffusion direction is selected based on the type of edge mode.
In addition, it is found that the edge modes are immune to defects.
The robust modes stably lead to directional heat transfer in realistic diffusion systems. 
The topological diffusion systems provide a fruitful avenue for temperature controls, thermal managements, and control for other diffusion phenomena with the topological insulators.

Again, our approach taken here exploits the topological edge modes in the honeycomb structure for the directional heat transfer.
This can potentially be extended to other diffusion phenomena, e.g., ion transport.
An interesting point is that the ionic transport is described by internal and external factors unlike the thermal diffusion. 
Specifically, ionic diffusivity is affected by the external factor (e.g., ambient temperature) as well as the internal material properties such as number of the mobile ions, charge, and activation energy. 
The ambient temperature can be adjusted by applying external signals. 
Hence, it will be possible actively to control the topological and ordinary states via the ambient temperature in ion transport systems.
Design for topological systems depending on both internal and external factors is a crucial future work.

\section{Methods}
\subsection{Numerical simulation of time evolving temperature}
The time evolution of the temperature is calculated by the MEMS module in COMSOL Multiphysics. 
We consider the finite structure consisting of 13 $\times$ 12 unit cells to obtain the edge states between the ordinary and topological unit cells, see Fig.~\ref{fig:figure3}(a). 
Each unit cell has six disc-shaped sites (thickness of 10 mm and diameter of 20 mm). 
The cells have intra-(inter-)cell connection using the beam of $L$ = 35 mm with the thermal diffusivity $D_1$ ($D_2$), see Fig.~\ref{fig:figure1}(a). 
We set the initial temperature to the unit cells, corresponding to the edge modes calculated from the eigen functions of the effective diffusivity matrix. 
We apply adiabatic boundary on whole surface of the structure to eliminate the thermal radiation and convection.
In the numerical model, we design the thermal diffusivities of the beams by directly adjusting the value of thermal conductivity in the original material parameters. 
The temperature is calculated in the time duration from 0 to 100 s with the step of 1 s.

\section{Data availability}
We declare that the data supporting the findings of this study are availabel within the paper.

\section{Author contributions}
All authors contributed extensivelly to the work presented in this paper.
H.T. managed this project.
K.F. and H.T. carried out the theoretical analysis with the assist of A.M and J.H.
K.F. designed and calculated the numerical model.


\providecommand{\noopsort}[1]{}\providecommand{\singleletter}[1]{#1}%

\end{document}